\documentclass[prd,aps,twocolumn,showpacs,preprintnumbers,amsmath,amssymb,
nofootinbib]{revtex4}
\usepackage{graphicx}

\newcommand{\be}{\begin{equation}}
\newcommand{\ee}{\end{equation}}
\newcommand{\bea}{\begin{eqnarray}}
\newcommand{\eea}{\end{eqnarray}}
\newcommand{\eq}[2]{\begin{equation}\label{#1} #2 \end{equation}} 
\newcommand{\N}{\mathcal{N}}
\newcommand{\alp}{\alpha'}

\newcommand{\del}{\partial}
\newcommand{\comment}[1]{}
\newcommand{\Z}{\mathbb{Z}_3}
\newcommand{\bi}{\bar{\imath}}
\newcommand{\bj}{\bar{\jmath}}
\newcommand{\NS}{\mathnormal{NS}}
\newcommand{\tNS}{\widetilde{\NS}}
\renewcommand{\t}[1]{\tilde{#1}}
\newcommand{\ket}[1]{| #1 \rangle}
\newcommand{\m}[1]{\mathbf{#1}}
\renewcommand{\b}[1]{\bar{#1}}

\begin{document}

\date{\today}
\title{Stabilizing moduli with thermal matter and nonperturbative effects}

\author{Rebecca J. Danos}\email[email: ]{rjdanos@hep.physics.mcgill.ca}
\affiliation{Department of Physics, McGill University, 
Montr\'eal, QC, H3A 2T8, Canada} 
\author{Andrew R. Frey}\email[email: ]{frey@hep.physics.mcgill.ca}
\affiliation{Department of Physics, McGill University, 
Montr\'eal, QC, H3A 2T8, Canada} 
\author{Robert H. Brandenberger}\email[email: ]{rhb@hep.physics.mcgill.ca}
\affiliation{Department of Physics, McGill University, 
Montr\'eal, QC, H3A 2T8, Canada} 


\begin{abstract}
Even with recent progress, it is still very much an open question to 
understand how all compactification moduli are stabilized, since there are
several mechanisms.  For example, it is possible to generate a scalar 
potential either classically or through nonperturbative effects, such as
gaugino condensation. Such a potential can stabilize
certain of the moduli fields, for example the dilaton. On the other hand, 
a background of thermal matter with moduli-dependent masses can also 
stabilize certain of the moduli, \textit{e.g.}, 
the radion.  It is important to 
understand whether these two distinct mechanisms are compatible with each
other, that is, that there are no interference terms that could spoil
the moduli stabilization.  In this paper, we study heterotic string theory
on an $\N=1$ orbifold near an enhanced symmetry point.  We then consider both 
a nonperturbatively generated potential and a gas of strings with 
moduli-dependent masses to stabilize the dilaton and radial modulus, 
respectively.  We conclude that, given certain approximations, these
two moduli stabilization mechanisms are compatible.
\end{abstract}
\maketitle

\section{Introduction}\label{s:intro}

One of the outstanding challenges in connecting string theory to cosmology
is to stabilize the many moduli fields predicted by string theory which are
not observed in our low energy world. These moduli fields include the
volume and shape moduli of the compact internal space, and the axion-dilaton
multiplet.

In the context of low energy supergravity limits of superstring theory, there
has over the past years been a large body of work devoted to stabilizing
these moduli fields. The most popular approach is to introduce 
fluxes about the compact spaces to stabilize the shape 
(``complex structure") moduli \cite{DRS,GKP} and to invoke non-perturbative
effects such as gluino condensation to stabilize the volume (``K\"ahler")
moduli \cite{KKLT} (see, \textit{e.g.}, \cite{ES,hep-th/0308156} 
for pedagogical overviews of
moduli stabilization in flux compactification scenarios). 
While these techniques are well-understood in type II string theory, the
situation in heterotic string theory is more murky, largely because 
the analogue of supergravity flux in the heterotic theory is a deformation of
the geometry away from Calabi-Yau manifolds 
\cite{Strominger:1986uh,deWit:1986xg,Hull:1985zy,Hull:1986kz}.  Indeed,
it is possible to fix all the moduli in type II compactifications by 
deforming the geometry and turning on general fluxes, but this approach is
not yet fully understood.

An alternative approach to moduli stabilization arises in string gas cosmology
\cite{BV} (for overviews see,\textit{e.g.}, \cite{RB1,BW,RB2,RB3} and
for some criticisms of the original scenario see 
\cite{Cleaver:1994bw,Sakellariadou:1995vk,Easther:2004sd,Danos:2004jz}). String
gas cosmology is based on coupling a gas of strings to a cosmological
background in the same way that Standard Cosmology arises from coupling
a gas of point particles to a background described by Einstein gravity.
At late times, the background for string gas cosmology can be taken
to be dilaton gravity \cite{TV,BEK}.\footnote{However, the dilaton
gravity background is inadequate to describe the early Hagedorn phase
of a string gas \cite{Betal,KKLM}. In particular, the recently
proposed string gas structure formation scenario \cite{NBV,BNPV2},
according to which thermal fluctuations of the string gas in the
Hagedorn phase lead to an approximately scale-invariant spectrum of
cosmological perturbations with a slight blue tilt \cite{BNPV1} to
the spectrum of gravity waves, requires a background which goes
beyond dilaton gravity. For attempts to construct such a background
see, \textit{e.g.}, \cite{BFK1}.}

In the context of string gas cosmology based on heterotic string
theory there is a geometric and specifically stringy mechanism
which stabilized both the volume moduli 
\cite{Watson1,Patil1,Patil2,Watson2,Rador2,Chatrabhuti,Kim,Easson:2005ug}
and also the shape moduli \cite{Edna} (see also \cite{Kaya2}).
However, this mechanism is unable to fix the dilaton modulus \cite{Biswas}
(for attempts to stabilize the dilaton see, \textit{e.g.}, 
\cite{Patil3,Sera,Battefeld,Kanno,Rador,Kaya}).
The key point is that there are states, in our case containing both momenta and
windings, which are massless at certain 
enhanced symmetry points in moduli space
\cite{Watson2,beauty}. The winding provides a force counteracting
expansion, while the momenta yield a force opposing the contraction. Together,
this yields an effective potential for the radion with a minimum
at a finite radius. Since the states are massless at this radius, the
magnitude of the potential vanishes at the enhanced symmetry
point, leading to radion stabilization. The existence of perturbative
states that
become massless at special values of the compactification radius is a 
special feature of heterotic string theory not shared by Type II theories 
(more elaborate constructions are necessary; for some examples, see
\cite{Strominger:1995cz,Polchinski:1995df,Katz:1996ht}).

In this paper, we consider a combination of these two mechanisms in a 
toy model based on a compactification of the heterotic string.
Specifically, we will consider heterotic
string theory on a $T^6/\Z$ orbifold, a textbook example of $\N=1$
string compactifications.  We will make use of the non-perturbative mechanism 
of gaugino condensation
\cite{Ferrara:1982qs,Affleck:1983rr,Affleck:1983mk,Affleck:1984xz,
Shifman:1987ia,Dine:1985rz} to stabilize the dilaton and Kaluza-Klein 
momentum-winding states that become massless at an enhanced symmetry point
to stabilize the radion.  The key question to address
in this context is whether this dilaton stabilization mechanism is
compatible with radion stabilization. Our preliminary results, based on
some simplifying approximations, are that
the answer to this question is positive.

In the following section we will review
the construction of this orbifold and discuss the spectrum of
massless states of the model. In Section III we construct the
potential for the radion/dilaton system which results from
gaugino condensation. Taken by itself, the resulting
potential can easily stabilize the dilaton. In Section IV 
we review the effective potential for the radion which results from a gas 
of massless string states. This potential rather trivially stabilizes
the string frame radion. In Section V we then combine both potentials
and demonstrate that both the radion and the dilaton are stabilized.
We conclude with a discussion of the stability of the enhanced symmetry
states, and give some general conclusions. 

\section{Heterotic on $T^6/\Z$}\label{s:orbifold}

The $T^6/\Z$ orbifold is one of the longest-studied compactifications of 
string theory, dating to the earliest days of the heterotic string
\cite{Dixon:1985jw,Dixon:1986jc,Witten:1985xb,Ferrara:1986qn}, 
and it is in fact a textbook example of $\N=1$ string
compactifications \cite{Polchinski:1998rr}.  In this section, we review
the construction of the orbifold.  
Then we describe the light degrees of freedom living on the orbifold 
near an enhanced symmetry point, including the moduli space.  We also 
discuss the assumptions we make to simplify our calculations.  We give a 
detailed derivation of the spectrum of light strings on the orbifold in
the appendix.

A brief remark on notation: we take Greek indices $\mu,\nu$ to represent
the external (large) spacetime dimensions, Roman indices $m,n$ to represent
all the internal dimensions, and $i,j$ ($\bi,\bj$) to represent just the 
(anti)holomorphic internal dimensions.  In this section, we work in the 
10D string frame when describing the orbifold construction and the spectrum
except in formula (\ref{KKfinal});
the K\"ahler potentials and kinetic actions given in subsection 
\ref{s:modulispace} are for the 4D Einstein frame, as usual, with the 4D
moduli defined in terms of the 10D string frame variables.

\subsection{Construction of the orbifold}\label{s:identification}

We begin with a six-torus with moduli chosen so that it factorizes into three
two-tori, $T^6=T^2\times T^2\times T^2$.  We make this choice not only to
allow the orbifold projection but also for clarity of presentation; we will
later discuss the moduli space on the orbifold.
With this factorization, we can
choose to coordinatize the $T^6$ with one complex coordinate $Z^i$ for each
$T^2$, and we take the coordinate periodicities to be the same on each
torus:
\eq{periods}{Z\simeq Z +2\pi\sqrt{\alp}\ , \ \ Z\simeq Z+2\pi\alpha\sqrt{\alp}
\ ,\ \ \alpha =e^{2\pi i/3} \ .}
In addition, we consider identical metrics on the $T^2$s,
\eq{T2metrics}{ds_{T^2}^2 = b^2 \left|dZ\right|^2\ .}
Here, $b$ is the scale factor of the metric, which controls the physical 
radius of the $T^2$.  Unlike other moduli, we will allow this modulus to vary,
as it illustrates the dependence of the string spectrum on the moduli in a 
simple manner. Considering identical metrics on the three different
tori is reasonable since, as shown in \cite{Watson0}, string gases
lead to an isotropization of an initially anisotropic space.

\begin{figure}
\includegraphics[scale=0.6]{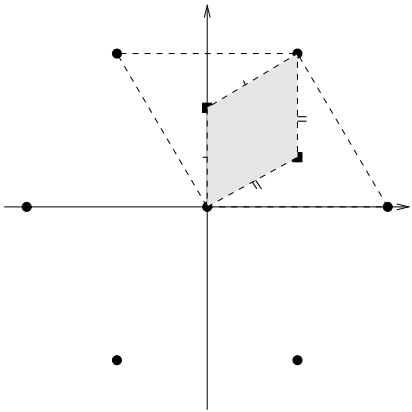}
\caption{The unit cell of a single $T^2/\Z$ factor of the $T^6/\Z$ orbifold.
Circular dots are the lattice points of the torus and square dots are the
additional fixed points of the orbifold.  The unshaded parallelogram is the
unit cell of the torus, and the shaded region is the unit cell of the orbifold.
Cross-lines indicate that these boundaries are glued together in the 
orbifold.\label{f:orbifold}}
\end{figure}

We can now quotient the $T^6$ by a $\Z$ symmetry, which acts on the three
complex coordinates as
\eq{Z3one}{Z^1\simeq \alpha Z^1\ , \ \ Z^2\simeq \alpha Z^2\ ,\ \ 
Z^3\simeq \alpha^{-2} Z^3\ .}
The resulting orbifold has $SU(3)$ holonomy and preserves $\N=1$ supersymmetry
\cite{Dixon:1985jw,Dixon:1986jc,Witten:1985xb}.  The geometric structure of 
this orbifold, including identifications and the unit cell, are shown in
figure \ref{f:orbifold}.

We also need to specify
the action of the $\Z$ group on the gauge degrees of freedom of the heterotic
theory; for simplicity, we take trivial Wilson lines for the gauge fields
(that is, we set the constant part of the gauge fields to zero, $A_m=0$) 
and embed the spin connection in the gauge connection (that is, we choose
a $\Z$ subgroup of the gauge theory and project onto states invariant under
the diagonal combination of both the gauge and geometrical $\Z$s).
This is the simplest possible model, as is illustrated in 
\cite{Polchinski:1998rr}.  (It is possible to design models with more realistic
spectra by taking nontrivial Wilson lines \cite{Font:1989aj}.)  Quotienting out
by this gauge group factor breaks the 10D gauge theory.  For example, 
in the $E_8\times E_8$ string, we are left with an unbroken 
$SU(3)\times E_6\times E_8$,
and the Standard Model is taken to live within the $E_6$ factor.

\subsection{Light degrees of freedom}\label{s:modulispace}

We will now describe the light degrees of freedom visible near a particular
enhanced symmetry point of the orbifold moduli space.  We start by describing
the moduli space itself.  For a detailed description of which states 
survive the orbifold projection, see the appendix.

The moduli space of the orbifold contains fields from both the untwisted and
twisted sectors; however, for reasons we will explain below, we will set 
all the twisted sector moduli to zero for the remainder of this paper.  The
remaining moduli space, with only untwisted modes, was first described by
\cite{Ferrara:1986qn} as a product of coset spaces
\eq{cosets}{\frac{SU(1,1)}{U(1)}\times \frac{U(3,3+81)}{U(3)\times U(3+81)}
\ ,}
where $81$ is a number of scalars from the gauge theory of the heterotic 
compactification.

The $SU(1,1)/U(1)$ factor describes the dilaton $\Phi$
and universal axion $a$ of the 
4D effective theory.  This axion is the 4D Hodge dual of the external 
2-form $B_{\mu\nu}$, $da=\star dB$, and the 4D dilaton is
$\Phi=2\phi-6\ln b$ in 10D string frame variables.  
Working in the 4D Einstein frame, 
these moduli form a complex scalar $S$ with K\"ahler potential
\eq{SKahler}{\mathcal{K}(S)=-\ln(S+\b S)\ , \ \ S=e^{-\Phi}+ia\ .}
Although we will not need it, we give here for reference the kinetic action
\bea
S &=& -M_P^2 \int d^4x\,\sqrt{-g_E} \frac{1}{(S+\b S)^2} \del_\mu S
\del^\mu \b S\nonumber\\
&=&\!\!\! -\frac{M_P^2}{4}\!\!\int\! d^4x\sqrt{-g_E} \left(
\del_\mu\Phi\del^\mu\Phi +e^{2\Phi}\del_\mu a\del^\mu a\right) .
\label{Skinetic}\eea

The other coset factor includes the internal metric and 2-form as well as the
internal gauge fields $A_i$ and $A_{\bi}$.  As the gauge degrees of freedom
are not important to our analysis (and have some constraints due to having
a nontrivial superpotential), we will assume that they vanish, as well.
We are now left with a $U(3,3)/U(3)\times U(3)$ coset to describe the 
moduli $T_{i\bj}\equiv g_{i\bj}+B_{i\bj}$.  These have K\"ahler potential
\eq{U3Kahler}{\mathcal{K}(T)=-\ln\det\left(T_{i\bj}+\b T_{i\bj}\right)}
and kinetic action
\bea S&=&-\frac{M_P^2}{4}\int d^4x \sqrt{-g_E}\, g^{\b l i} g^{\bj k}
\left( \del_\mu g_{i\bj}\del^\mu g_{k\b l}\right.\nonumber\\
&&\left. -\del_\mu B_{i\bj}\del^\mu B_{k\b l} \right)\ .\label{U3kinetic}\eea
The 2-form has the wrong-sign kinetic term because it is antihermitean.
(For an example of this coset moduli space in flux compactifications of 
the type IIB string, see \cite{hep-th/0201029}.)

For simplicity, we will now assume that $g_{i\bj}$ and $B_{i\bj}$ are
diagonal, so the $T^6/\Z$ factorizes into three $T^2$ factors.  We further
assume that the moduli are the idential for each $T^2$ factor.  In other
words,
\eq{simplified}{g_{i\bj} =\frac{b^2}{2}\delta_{i\bj} , \ B_{i\bj}=
i\frac{\beta}{2}\delta_{i\bj} , \ T_{i\bj}= \frac{T}{2}\delta_{i\bj} .}
(As we write it here, $T=b^2+i\beta$ 
is most conveniently described in terms of the 10D
string frame variables.)
Note that the complex structure (shape) moduli of the $T^2$ factors, which 
correspond to metric components $g_{ij},g_{\bi,\bj}$, have already been
removed by the orbifold projection.  The K\"ahler potential now becomes 
\eq{TKahler}{\mathcal{K}(T)=-3\ln(T+\b T)} (up to an irrelevant constant).

Next we consider the moduli space with the
twisted sectors included.  When some of the twisted sector moduli have 
nonzero expectation values, others become massive, and the orbifold fixed 
points ``blow up'' into smooth spaces.  Considering the total 
compactification, $\N=1$ supersymmetric orbifolds are deformed by 
their twisted sector moduli into Calabi-Yau manifolds
\cite{Hamidi:1986vh,Dixon:1986qv}, and the moduli space loses the coset
space form of (\ref{cosets}).  As a result, with nonvanishing twisted 
sector moduli, we would lose the ability to describe the string spectrum
explicitly, so we choose to study an ansatz with the twisted sector set to 
zero.  However, since the twisted sectors deform the orbifold into a 
non-flat space, 
the states (\ref{enhanced1},\ref{enhanced2}) below, which become
massless at the orbifold enhanced symmetry point, should also gain a mass from 
the twisted sector moduli, in which case they can stabilize the twisted 
sectors as well as the radius.  In addition, as we mention below 
(\ref{nonpertW2}), gaugino condensation can also stabilize complex structure
moduli, which provides additional support for this ansatz.  Finally,
we also do not expect our results to change much with the inclusion of the
twisted sector moduli because they are complex structure moduli and couple
to the K\"ahler moduli only in the potential. 
We do not consider this problem further in this paper. 

Finally, let us give a few words about the Kaluza-Klein (KK) momentum and
winding states of the orbifold, which are described in the appendix.  
As the orbifold approaches the point $T=1+i/\sqrt{3}$ in moduli space, 
there are a number of string states that become massless.  In particular, 
there are 6 new massless vectors (describing a $U(1)^6$ gauge factor), 18
complex scalars, and their supersymmetric partners.  From the perspective
of the 10D string theory, though, these are all just strings with single units
of KK winding and momentum.  If we hold $\beta$ fixed and let $b=1+\delta b$,
these states gain a mass $m^2\approx 4\delta b^2/\alp$, which is just the
same as the states considered in \cite{Patil2}.  Therefore, we may use the
string gas energy-momentum tensor derived in \cite{Patil2} to describe 
the gas of these momentum/winding strings very close to the enhanced symmetry
point.  We do work in the 10D Einstein frame, however, so we need to take
$b\to e^{\phi/4}b$.  As discussed in the appendix, we find
\eq{KKfinal}{
m^2 = \frac{1}{\alp} \left(e^{-\phi/4}b^{-1}-e^{\phi/4}b\right)^2 
\ .}  We will use this version from now on.

\section{Gaugino condensation and moduli stabilization}

It was recognized in the early 1980s that a nonzero expectation value for
a fermion bilinear would break supersymmetry in supergravity theories
\cite{Ferrara:1982qs}.  Shortly thereafter,
\cite{Affleck:1983rr,Affleck:1983mk,Affleck:1984xz,Shifman:1987ia} proved
that such expectation values do form in $SU(N)$ and $SO(N)$ gauge theories
with sufficiently few matter fields. It is extremely natural to extend these
results to $\N=1$ compactifications of string theory \cite{Dine:1985rz}, 
and the
supersymmetry breaking also leads to a potential for the compactification
moduli.

In this section, we will review the relevant physics of gaugino condensation in
supersymmetric field theories, as well as explain how it plays a role in
the stabilization of moduli in string compactifications.  Once we have
laid out the background, we will derive the potential for the moduli in
our simplified ansatz.  In this section, we work in the 4D Einstein frame
until stated otherwise.

\subsection{The nonperturbative superpotential}

One of the most basic facts about supersymmetric gauge theories is that
the gauge bosons have fermion superpartners, and these gaugini can
be combined into a gauge and Lorentz invariant bilinear.  As 
\cite{Affleck:1983rr,Affleck:1983mk,Affleck:1984xz,Shifman:1987ia} showed,
these bilinears quantum mechanically develop nonzero expectation values,
sometimes through physics as calculable as instanton effects
(under some conditions on other fields present in the theory); this 
expectation value is known as the gaugino condensate.

The important feature of gaugino condensation is that it induces a 
correction to the superpotential $W$ of the theory, which is exponentially 
suppressed by the gauge coupling $g$:
\eq{nonpertW1}{W\to W-Ae^{-1/g^2}} (the sign is chosen for later convenience).
More precisely, the $\theta$ angle of the gauge theory should be included,
which takes $1/g^2\to 1/g^2+i\theta/2\pi$ in (\ref{nonpertW1}).  
This combination of coupling and $\theta$ angle is known as the gauge
kinetic function; in 
supergravity constructions, including $\N=1$ string theory compactifications,
the gauge kinetic function is given precisely by the modulus $S$ defined
by (\ref{SKahler}) \cite{Ferrara:1982qs,Dine:1985rz}.  

In a Calabi-Yau compactification of the heterotic string (which, broadly 
defined, includes our orbifold as a special point in moduli space), 
the tree-level superpotential is given by the field strength $H$
\cite{Dine:1985rz}, which gets contributions from the derivative of 
$B_{mn}$ and from gauge theory Wilson lines (which could, for example,
break the $E_6$ gauge group to the Standard Model or some other grand unified
group or alternately break the $E_8$ factor to a smaller group) (for an 
analysis of Wilson lines in this context, see \cite{hep-th/0310159}).
The key feature of this superpotential, which we will call $W_H$, is its
independence of the volume modulus $T$ (for the moment, we ignore the
complex structure moduli).  Following the logic of \cite{Dine:1985rz},
the gaugino condensate and $H$-flux should cancel each other's effects on the
compactification, which leaves a nonzero value for the superpotential in
the ground state and also breaks supersymmetry.  Compactifications with
gaugino condensation and $H$-flux on non-Calabi-Yau manifolds were studied in
\cite{hep-th/0310021,hep-th/0507173,hep-th/0507202}.

We now have a total superpotential 
\eq{nonpertW2}{W = W_H - A'e^{-a_0S} \equiv M_P^3\left( 
C-A e^{-a_0S}\right)\ ,}
with $a_0,A,C$ some constants.  At first glance, and as we review below,
it seems that this superpotential can stabilize the dilaton modulus $S$ but
not the volume modulus $T$.  While it is not obvious from our discussion,
this superpotential can also stabilize the complex structure moduli of a 
Calabi-Yau manifold, due to the requirement that the gaugino condensate and
$H$-flux align in the extra dimensions \cite{Dine:1985rz}.  This fact provides
some justification for our assumption that the twisted sector moduli vanish;
we are in effect assuming that the gaugino condensate has already stabilized
these moduli.

One last comment before we move to the potential: the gaugino condensate and
$H$-flux are new ingredients to the string compactification and can 
potentially affect the spectrum of the compactification (for example, the 
superpotential (\ref{nonpertW2}) will give a mass to the dilaton).  
It is certainly possible (or even likely) that these nonperturbative 
ingredients give the special momentum/winding states we consider a small 
mass.  Nonetheless, we expect that the minimal mass is still near the 
original enhanced symmetry point, so moduli stabilization should still occur.
We further believe that the
correction to the mass formulae (\ref{KKfinal}) should be small, since they are
due to nonperturbative effects.  A related issue is that we do not have the
means to study interactions between the gaugino condensate and momentum/winding
strings from the 10D perspective (for a few beginning steps in this direction,
see \cite{hep-th/0507202,hep-th/0509082,arXiv:0707.1038}).  A complete
answer to these questions must wait for a detailed study of the 4D effective
field theory, which we leave for future work.

\subsection{Complete potential}

In the 4D Einstein frame, the scalar potential is given in terms of the 
superpotential and K\"ahler potential by the well-known formula
\eq{scalarV1}{V=\frac{1}{M_P^2}e^{\mathcal{K}}\left(\mathcal{K}^{AB}
D_A W D_{\b B}\b W -3|W|^2\right)\ ,}
where $A,B$ run over all moduli and 
the K\"ahler covariant derivatives are given by
\eq{KahlerD}{D_A W = \del_A W + (\del_A \mathcal{K})W\ .}
Because the superpotential is independent of the volume modulus $T$,
(\ref{scalarV1}) simplifies to
\eq{scalarV2}{V = \frac{1}{M_P^2}e^{\mathcal{K}}\mathcal{K}^{ab}
D_a W D_{\b b}\b W\ ,}
where $a,b$ now run over $S$ (and the complex structure moduli, if we wish to
consider them).  In fact, if we allow more general moduli $T_{i\bj}$, 
all of them drop out, leaving the same formula (\ref{scalarV2}).

We can immediately make a few important statements about the potential based
on the so-called ``no-scale'' structure of (\ref{scalarV2}).  First is
that $V$ is the sum of squares, so it is minimized precisely when each term 
(in our case, there will be only one term) vanishes.  This means that the
cosmological constant from this potential will vanish.  Additionally, 
$V$ depends on the $T$ modulus only through the overall factor of 
$e^{\mathcal{K}}$, so $T$ is not stabilized classically in this potential.
(See \cite{hep-th/0310159} for a discussion of quantum mechanical corrections
and their effects on $T$.)  We will be interested in using string matter
to stabilize $T$, however, so we will not consider loop effects.

In our case, after we apply all the simplifications to the moduli space,
we get
\bea
V&=&\frac{M_P^4}{4} b^{-6} e^{-\Phi}\left[\frac{C^2}{4}e^{2\Phi} +ACe^\Phi
\left(a_0+\frac{1}{2}e^\Phi\right)e^{-a_0e^{-\Phi}}\right.\nonumber\\
&&\left. +A^2 \left(a_0+\frac{1}{2}e^\Phi\right)^2e^{-2a_0e^{-\Phi}}\right]\ .
\label{scalarV3}\eea
For our purposes, it is sufficient to expand around a minimum in $\Phi$, 
so we can approximate the potential as
\bea V&=&\frac{M_P^4}{4} b^{-6} e^{-\Phi_0} a_0^2 A^2 \left(a_0-\frac{3}{2}
e^{\Phi_0}\right)^2 e^{-2a_0 e^{-\Phi_0}}\nonumber\\ 
&&\times\left(e^{-\Phi}-e^{-\Phi_0}\right)^2\ .\label{scalarV4}\eea

Finally, we will be working in the 10D Einstein frame, so we need to lift
this potential to that frame.  In the 10D Einstein frame, the 
metric is
\eq{ds10D}{ds^2 = b_E^{-6} g^E_{\mu\nu}dx^\mu dx^\nu + 
b_E^2 \hat g_{mn}dx^m dx^n
} with 4D Einstein frame metric $g^E_{\mu\nu}$ and a fixed fiducial metric
$\hat g_{mn}$ on the compact space.  The Planck mass is therefore given 
by $M_P^2 = M_{10}^8\hat V$ in terms of the 10D Planck mass and the fiducial
volume of the compactification.  Furthermore, the potential term in the 
action is
\eq{uplift1}{\int d^4x \sqrt{-g^E} V_E =\frac{1}{\hat V} \int d^{10}x
\sqrt{-g}\, b_E^6 V_E\ ,}
which implies
\eq{uplift2}{V=\frac{b_E^6 V_E}{\hat V}\ .}
Using the relation of the 4D to 10D dilaton and the relation of the string and
Einstein frame scale factors $b$, we find
\bea \label{gaugepot}
V(b,\phi) &=& \frac{M_{10}^{16}\hat V}{4} e^{-\Phi_0}
a_0^2 A^2 \left(a_0-\frac{3}{2}e^{\Phi_0}\right)^2
e^{-2a_0 e^{-\Phi_0}}\nonumber\\
&&\times e^{-3\phi/2}\left(b^6 e^{-\phi/2}-e^{-\Phi_0}\right)^2\ .
\label{Vfinal}\eea
We have now written the scale factor in the 10D Einstein frame.  Also, we
have written the potential so that the entire first line is just an overall
constant.

\section{Moduli stabilization by matter}

In this section we review how for fixed dilaton a gas of
\textit{massless} string states leads to radion stablization. By
\textit{massless} strings we mean string states which become massless
at a special value of the radion. We will follow the discussion
in Section 3 of \cite{Patil2}.

Assuming homogeneity and isotropy of our three spatial dimensions, the
metric of our ten-dimensional space-time can be written as
\be \label{metric}
ds^2  =  - dt^2 + a(t)^2 d\vec{x}^2 + 
\sum_{i = 1}^3 b(t)^2 |dZ_i|^2 \, ,
\ee
where $a(t)$ is the scale factor of our three dimensions, and 
$b(t)$ is the scale factor of the internal space. To
simplify the analysis, we set the scale factors of all three
internal two-tori equal.

The total action of our system is given by the action of dilaton
gravity coupled to a gas of string states. More specifically,
the action is
\be \label{totaction}
S =  \frac{1}{\kappa}\left( S_g + S_{\phi}\right) + S_{SG} \, ,
\ee
the first term standing for the usual Einstein action, the second
term denoting the dilaton action and the third the action of the
string gas. In the above,
\be
\kappa =  16 \pi G =2/M_{10}^8
\ee
is given by the 10-dimensional gravitational constant $G$ or equivalently
the 10D Planck mass $M_{10}$.

We will be working in the ten-dimensional Einstein frame. Thus,
the dilaton action is
\be
S_{\phi}  = -\int d^{10}x\sqrt{-g}\left[\frac{1}{2}
\partial_{\mu}\phi\partial^{\mu}\phi + \kappa V(\phi)\right] \, ,
\ee
where $V(\phi)$ is the dilaton potential discussed in the
previous section. 
In this section, we will assume that the dilaton is fixed by hand
and review how in this setup a gas of special string states will
lead to radion stabilization. 

Treating the strings in the ideal gas approximation, the action for
a string gas can be written by summing over the contributions to the
gas from each string state, labelled below by $\alpha$. The individual
contribution is obtained from its number density 
$\mu_{\alpha}(\vec x, t)$ and energy $\epsilon_{\alpha}(t)$ in a hydrodynamic
approximation:
\be \label{SGaction1}
S_{SG}  =  -\int d^{10}x\sqrt{-g} \sum_{\alpha} \mu_{\alpha}
\epsilon_{\alpha} \, .
\ee
In the homogeneous approximation the densities are independent of
$\vec{x}$. Thus, the spatial metric can be factored out of the
number density via
\be
\mu_{\alpha} \, = \, \frac{\mu_{0, \alpha}(t)}{\sqrt{g_s}} \, ,
\ee
where $g_s$ is the determinant of the spatial part of the metric.
Thus, (\ref{SGaction1}) becomes
\be
S_{SG} = -\int d^{10}x\sqrt{-g_{00}} 
\sum_{\alpha} \mu_0(t)_{\alpha} \epsilon_{\alpha} \, .
\ee

From the above, we can derive the components of the string gas
energy-momentum tensor which will enter the cosmological equations
of motion, namely the energy density
\be
\label{tt}
\rho_{\alpha} \, = \, 
\frac{\mu_{0, \alpha}}{\epsilon_{\alpha}\sqrt{-g}} \epsilon^2_{\alpha} \, ,
\ee
the pressure in our three dimensions
\be
\label{ii}
p^i_{\alpha} \, = \,  
\frac{\mu_{0, \alpha}}{\epsilon_{\alpha}\sqrt{-g}}\frac{p^2_{d}}{3} \, ,
\ee
where $p_{d}$ is the momentum in our $d=3$ large dimensions,
and the pressure in the compact directions
\be
\label{aa}
p^a_{\alpha} \, = \,  
\frac{\mu_{0, \alpha}}{\epsilon_{\alpha}\sqrt{-g}\alpha'}
\left( \frac{n_a^2}{b_a^2} - w_a^2 b_a^2 \right) \, .
\ee    
In the above, we have
focused on the contributions of a particular string state $\alpha$
with energy $\epsilon_{\alpha}$
\bea 
\epsilon_\alpha
&=&
\frac{1}{\sqrt{\alpha'}}\left[\alpha' p^2_{d} + b^{-2} (n,n) + b^2(w,w) 
\right.\nonumber\\
&&\left. +2(n,w) + 4(N-1)\right]^{1/2} 
\, ,\label{mass}
\eea
where $\vec{n}$ and $\vec{w}$ are the momentum and winding number
vectors in the internal space, $\vec{p}$ is the momentum in
the non-compact space (indicated with subscripts $d$ for the $d$ large 
dimensions), and
$N$ is the oscillator level. The parentheses in the momentum
and winding number terms on the right hand side of the above
equation indicate scalar products
in the internal space, with the modulus field $b(t)$ factored out.  From 
now on, we assume that the internal space is isotropic, so that all $b_a=b$.
 
The Einstein equations for the metric (\ref{metric}) which follow from
the above expressions for the energy-momentum tensor of the string gas
composed of strings with fixed quantum numbers $n_a$ (momenta), $w_a$
(windings) and $N$ (oscillator level) are 
\begin{eqnarray}
\label{c}
\ddot{b}\!\! &+&\!\! \dot{b}(3 \frac{\dot{a}}{a} 
+ 5 \frac{\dot{b}}{b}) \,
= \, \frac{8\pi G \mu_{0, \alpha}}{\alpha'\sqrt{\hat{G}_a}
\epsilon_{\alpha}} \\
&\times &  \left[ \frac{n^2_a}{b^2} - w_a^2 b^2 
+  \frac{2}{9}[b^2(w,w) + (n,w) + 2(N-1)] \right] \nonumber \\
\label{nc}
\ddot{a} &+& \dot{a} (2 \frac{\dot{a}}{a} + 6 \frac{\dot{b}}{b}) \,
= \, \frac{8\pi G \mu_{0,\alpha}}{\sqrt{\hat{G}_i}
\epsilon_{\alpha}} \\
&\times& \left[ \frac{p^2_{d}}{3} + \frac{2}{9\alpha'}
[b^2(w,w) + (n,w) + 2(N-1)] \right] \, , \nonumber
\end{eqnarray} 
where $\hat{G}_\mu$ is the determinant of the metric without the 
$\mu$'th diagonal element, the subscript $a$ on the momentum and winding
numbers $n_a$ and $w_a$ label the direction
of the momentum and winding in the compact small dimensions and are to be
summed over.  
 
For the special enhanced symmetry states discussed in the appendix,
then, at the radius $b$ at which the enhanced symmetry appears, the
expression inside the square parentheses in (\ref{c}) and (\ref{nc})
vanish, \textit{i.e.},
\be
b^2(w,w) + (n,w) + 2(N-1) \, = \, 0 \, .
\ee
Looking at the equation (\ref{c}) for the radion, we see that the winding
numbers lead to a force resisting expansion, the momentum numbers lead
to a force resisting contraction. There is a stable fixed point at which
\be
\frac{n^2_a}{b^2} - w_a^2 b^2 \, = \, 0 \, .
\ee
Moving on to the equation (\ref{nc}) for the scale factor of our three
large spatial dimensions, we see that enhanced symmetry states at
the distinguished value of the radion act as radiation from the
effective four space-time dimensional point of view.

As already stressed in \cite{Patil2}, it follows from (\ref{c}) 
that unwound strings with $n_a = w_a=0$ at 
the oscillator level $N=1$ do not contribute to the driving term 
of the equation of motion for the compact dimensions. 
These states correspond to gravitons. 
In fact, any matter which is pressureless 
along the compact dimensions (such as ordinary matter) and satisfies the 
equation of state ($p = \rho/3$) does not contribute to the driving 
term for the compact dimensions. 

We thus see that a string gas on an internal manifold which admits stable
(or long-lived) winding modes automatically leads to radion stabilization
provided that the dilaton is held fixed. A similar dynamical analysis
\cite{Edna} shows that the shape moduli can also be stabilized. However,
if the dilaton is free to move according to the background equations of
motion for perturbative string theory, \textit{e.g.} dilaton gravity, the
dilaton is not stable and runs off to a singularity 
\cite{Biswas,Betal}, with fixed string frame 
radion \cite{Patil1}. We will now show that if we include gaugino
condensation, it is possible to fix simultaneously the dilaton and the
radion.

\section{Moduli stabilization with both effects}

\subsection{Combined equations of motion in 10D}

The energy-momentum tensor discussed in the previous section was in
terms of string frame quantitites (which coincide with Einstein frame
quantities if the dilaton is fixed). In this section the dilaton is
not fixed. We are working in the Einstein frame, and hence must convert
the string frame quantities (subscript $s$) 
to Einstein frame quantities
(subscript $E$) via the relations
\bea
g^E_{\mu\nu} \, &=& \, e^{-\phi/2}g^s_{\mu\nu} \\
b_s \, &=& \, e^{\phi/4}b_E \\
T_{\mu\nu}^E \, &=& \, e^{2\phi}T_{\mu\nu}^s \, .
\eea
The scalar products in the internal space used in (\ref{c}) and (\ref{nc})
also are modified according to
\bea
b^{-2}(n,n) \, &\rightarrow& \, e^{-\phi/2}b^{-2}n^2\\
b^2(w,w) \, &\rightarrow& \, e^{\phi/2}b^2w^2\\
(n,w) \, &\rightarrow& \, n\cdot w \, .
\eea

The dilaton potential which arises from gaugino condensation was derived 
in Section III and takes the form
\be \label{potential}
V \simeq n_1e^{-3\phi/2}\left(b^6e^{-\phi/2} - n_2\right)^2
\ee
where the coefficient $n_1$ can be read off from (\ref{gaugepot}) and 
$n_2=e^{-\Phi_0}$.

By varying the total action (\ref{totaction}) with respect to $\phi$, 
one can derive the following equation of motion for the dilaton
\bea\label{dilatoneom} 
&-& \frac{M_{10}^8}{2}\left(3a^2\dot{a}b^6\dot{\phi} 
+ 6a^3b^5\dot{b}\dot{\phi} +  
a^3b^6\ddot{\phi}\right) \nonumber \\
&+& \frac{3}{2}n_1a^3b^6e^{-3\phi/2}\left(b^6e^{-\phi/2}-n_2\right)^2  
\nonumber \\ 
&+& a^3b^{12}n_1e^{-2\phi}\left(b^6e^{-\phi/2}-n_2\right) \nonumber \\
&+& \frac{1}{2\epsilon}e^{\phi/4}\left(-\mu_0\epsilon^2 +
\mu_0|p_d|^2\right.  \nonumber \\
&+ & \left.\, 6 \mu_0\left[\frac{n_a^2}{\alpha^{'}}e^{-\phi/2}b^{-2}
-\frac{w^2}{\alpha^{'}}e^{\phi/2}b^2\right]\right) \nonumber 
\\
&=& \, 0 \, ,
\eea
where the energy $\epsilon$ of the string state is now given by
\bea
\epsilon  \, &=& \, \left(|p_{d}|^2 + 
\frac{e^{-\phi/2}b^{-2}n^2}{\alpha^{'}}\right. \\ 
&+&\left. \frac{1}{\alpha^{'}}e^{\phi/2}b^2w^2 + 
\frac{1}{\alpha^{'}}[2n\cdot w + 4(N-1)]\right)^{1/2} \, . \nonumber
\eea

For enhanced symmetry states, then at the enhanced symmetry value for
the radion $b$, the expression in the second to last line of 
(\ref{dilatoneom}) vanishes, and the contributions from the pressure
and energy density in the previous line cancel. Thus, if the
dilaton sits at the minimum of its potential, \textit{i.e.},
\be \label{enhanced}
b^6e^{-\phi/2}-n_2 \, = \, 0 \, ,
\ee
then the dilaton remains at rest. This value of the dilaton is a fixed
point of the dilaton equation of motion.

By inserting the energy-momentum tensor of both the dilaton and the
string gas into the trace-reversed Einstein equations, the internal
components of this equation yield the equation of motion for the
radion, which reads
\bea \label{radioneq}
&\ddot{b}& + 3\frac{\dot{a}}{a}\dot{b} + 5\frac{\dot{b}^2}{b} \\ 
&=& \, -\frac{n_1b}{M_{10}^8}e^{-3\phi/2}\left(b^6e^{-\phi/2}-n_2
\right)^2 \nonumber \\
&&-\frac{2n_1}{M_{10}^8} b^7e^{-2\phi}\left(b^6 e^{-\phi/2}-n_2\right) 
\nonumber \\
&&-\frac{1}{8}\left[-\frac{10b}{M_{10}^8}n_1e^{-3\phi/2}
\left(b^6e^{-\phi/2}-n_2\right)^2\right.
\nonumber \\
&&\left.-\frac{12n_1}{M_{10}^8}b^7e^{-2\phi}\left(b^6 e^{-\phi/2}-n_2
\right)\right] \nonumber \\
&&+\frac{8\pi G\mu_{0}}{\alpha^{'}\sqrt{\hat{G}_a}\epsilon}e^{2\phi}
\left[n_a^2b^{-2}e^{-\phi/2}-w_a^2b^2e^{\phi/2}  \vphantom{\frac{2}{D}}\right.
\nonumber \\
&&\left.+\frac{2}{9}(e^{\phi/2}b^2w^2 + n\cdot w + 2(N-1))\right] \nonumber
\eea

If the dilaton sits at the bottom of its potential and the radion takes on
the value given by (\ref{enhanced}), then the right hand side of
(\ref{radioneq}) vanishes. This value of the radion is hence a fixed
point of the radion equation.

We have now seen that a fixed point of the dynamical system is achieved
if the dilaton sits at the bottom of its potential and the radion takes
on the enhanced symmetry value (\ref{enhanced}). In the following, we
will demonstrate that this fixed point is stable. This will then complete
the demonstration that the joint action of gaugino condensation and
of a string gas consisting of enhanced symmetry states can stabilize both
the dilaton and the radion.

To demonstrate the stability of the above fixed point, we will linearize
the equations of motion (\ref{dilatoneom}) and (\ref{radioneq}) about the
fixed point and study their stability. After substituting 
$b=b_0 + \delta b$ and $\phi=\phi_0+\delta\phi$, where $b_0$ and $\phi_0$
form the fixed point solution, and keeping only linear terms in
the fluctuation variables $\delta b$ and $\delta \phi$, the linearized 
dilaton takes the form
\bea \label{lindil}
0 \, &=& \, \delta\ddot{\phi} + 3\frac{\dot{a}}{a}\delta\dot{\phi} \\
&& - \frac{2}{M_{10}^8}\left(6b_0^{11} e^{-5\phi_0/2}n_1
-\frac{12}{a^3b_0^5\alpha'}\frac{e^{3\phi_0/4}}{|p_d|}w^2\mu_0\right)\delta b
\nonumber \\
&& + \frac{2}{M_{10}^8}\left(
\frac{1}{2}b_0^{12} e^{-5\phi_0/2}n_1+\frac{3}{|p_d|\alpha'}
\frac{1}{a^3b_0^4}e^{3\phi_0/4}w^2\mu_0\right)\delta\phi \, ,\nonumber 
\eea
and the linearized radion equation is
\bea \label{linrad}
0 \, &=& \, \delta\ddot{b} + 3\frac{\dot{a}}{a}\delta\dot{b} \\
&& + \left(\frac{3b_0^{12}}{M_{10}^8} e^{-5\phi_0/2} n_1+\frac{32}{9|p_d|}b_0
e^{5\phi_0/2}n_3w^2\right)\delta b \nonumber \\
&& + \left(\frac{8}{9|p_d|}b_0^2 e^{5\phi_0/2}n_3 w^2
-\frac{b_0^{13}e^{-5\phi_0/2}n_1}{4M_{10}^8}\right)\delta\phi\ , \nonumber
\eea
where 
\be
n_3 \, = \, \frac{8\pi G\mu_{0}}{\alpha^{'}\sqrt{\hat{G}_a}} \, .
\ee

\subsection{Stability analysis of the equations}

The linearized equations (\ref{lindil}) and (\ref{linrad}) take the
form
\bea \label{linearizedeom1}
\delta\ddot{b} + 3\frac{\dot{a}}{a}\delta\dot{b}+A\delta b + B\delta\phi \, 
&=& \, 0\\
\label{linearizedeom2}
\delta\ddot{\phi} + 3\frac{\dot{a}}{a}\delta\dot{\phi}+C\delta b + 
D\delta\phi \, &=& \, 0 \, ,
\eea
where the coefficients $A, B, C$ and $D$ can be read off from
(\ref{lindil}) and (\ref{linrad}). 

To demonstrate the
stability of the fixed point, we need to show that the eigenvalues
of this system of differential equations are positive semidefinite.
We can neglect the Hubble damping terms since their effect is to
stabilize the dynamics rather than to destabilize it. 
Neglecting the Hubble damping terms, the system of equations is
that of two coupled harmonic oscillators. 
To find the eigenvalues, we make the eigenfunction ansatz
\be
\left[\begin{array}{c}
\delta b\\
\delta\phi\end{array}\right] = e^{i\omega t}\left[\begin{array}{c}
\delta b_0\\
\delta\phi_0\end{array}\right] \, .
\ee
The eigenvalues $\omega$ are then given by
\be
\omega^2 \, = \,  \frac{1}{2}\left[(D+A)\pm\sqrt{(D+A)^2-4(AD-BC)}\right]
\, .
\ee

The conditions for stability are that the values for $\omega^2$ are
real and positive. The reality condition translates to
\be \label{cond1}
(A - D)^2 + 4 BC \, \ge \, 0 \, ,
\ee
and the positivity condition is
\be
(D+A)^2 \, \ge \, (D+A)^2-4(AD-BC) \, ,
\ee
or, equivalently,
\be \label{cond2}
AD \, \ge \, BC \, .
\ee
The values for $AD$ and $BC$ are
\bea
AD \, &=& \, \frac{3b_0^{24}e^{-5\phi_0}n_1^2}{M_{10}^{16}}
+\frac{32 b_0^{13}n_1 n_3 w^2}{9|p_d|M_{10}^8} \\
&&+\frac{18b_0^8e^{-7\phi_0/4}n_1 w^2\mu_0}{a^3|p_d|\alpha' M_{10}^{16}}
+\frac{64e^{13\phi_0/4}n_3 w^4\mu_0}{3a^3 |p_d|^2\alpha' b_0^3 M_{10}^8}
\nonumber \\
BC \, &=& \, \frac{3b_0^{24}e^{-5\phi_0}n_1^2}{M_{10}^{16}}
-\frac{32 b_0^{13}n_1 n_3 w^2}{3|p_d|M_{10}^8} \\ 
&&-\frac{6b_0^8e^{-7\phi_0/4}n_1 w^2\mu_0}{a^3|p_d|\alpha' M_{10}^{16}} 
+\frac{64e^{13\phi_0/4}n_3 w^4\mu_0}{3a^3 |p_d|^2\alpha' b_0^3 M_{10}^8}
\nonumber 
\eea
By comparing the coefficients, it is trivial to see that the positivity
condition (\ref{cond2}) is satisfied. The reality condition
(\ref{cond1}) is automatically satisfied if $BC$ is positive. For
large $\mu_0$, it follows by direct inspection that both $B$ and $C$
are positive, and hence the reality condition holds. If $\mu_0$ does
not dominate the individual expressions for $B$ and $C$, we can
use another reasoning to argue that $BC > 0$ and hence the
reality condition is satisfied: it is again easy to see that both
in the limit $|p_d|\to\infty$ and in the limit $|p_d|\to 0$, the
expression $BC$ is positive.

To summarize, in this subsection we have shown, by considering the
linear fluctuation equations, that the fixed point at which the
dilaton sits in the minimum of its potential and the radion is
at the enhanced symmetry value is stable. This demonstrates that
both the radion and the dilaton can be simultaneously stabilized
by the combined action of enhanced symmetry string states and
gaugino condensation.

\subsection{Stability of KK states}

Now we must face one consequence of the orbifold compactification; strings
with momentum and winding around the torus do not generally carry any 
conserved charge and are therefore free to decay.  Semiclassically, we can 
think of the strings as ``unwinding'' around the fixed points of the
orbifold.  However, there are a few reasons that these decays are not a 
major concern.

First, except in the case of a scalar decaying into two scalars, the dominant
decay rate $\chi\to \psi\psi$ (for some generic fields) goes as 
$\Gamma \lesssim m$, where $m$ is the moduli-dependent mass of the $\chi$.
(There are of course suppressions by powers of the string coupling.)
In the linearized regime, this mass is very small, and 
we expect that the Hubble rate is faster.  Since the radion settles
at the enhanced symmetry point in about a Hubble time, the special 
momentum-winding states do not have time to decay before they become massless.
Also, because the effective theory on the orbifold has (spontaneously broken)
supersymmetry, the potential is a sum of squares.  If we try to generate a 
term $\chi\psi\psi'$, where $\psi$ and $\psi'$ are some moduli, we at the
same time introduce a field-independent mass for at least one of the moduli
fields.  Since these moduli are unstabilized unless there is a background of
$\chi$ particles, these terms to not exist.

A more important reason is parametric resonance, as explained in 
\cite{beauty,Watson2}.  As the radion passes near the enhanced symmetry point, 
the $\chi$ particles become very light and are produced via parametric
resonance.  Therefore, even if the $\chi$ particles decay, they will be
repopulated, which is an important point to remember about moduli stabilization
by matter.

\section{Discussion and Conclusions}

In this paper we have studied the compatibility of gaugino condensation
(which arises in $\N=1$ string compactifications and leads to 
dilaton stabilization) and
gases of enhanced symmetry string states (introduced in order to
stabilize the radion). We consider the action of dilaton gravity
coupled to a gas of heterotic 
strings treated in the ideal gas approximation,
in a space-time in which the internal space is the toroidal orbifold
$T^6/\Z$. We added to this system the potential for the dilaton
which emerges from gaugino condensation.  This is a simple but necessary 
check on the consistency of two types of moduli stabilization.

We investigated the
dynamical equations of motion for the dilaton and the radion which
follow from the full action of the system and identified a stable
fixed point which corresponds to the dilaton sitting at the minimum
of its potential and the radion taking on the value at which the
enhanced symmetry states are massless. The stability of this
fixed point was demonstrated by studying the linearized equations
of motion obtained by expanding about this fixed point and demonstrating
that the solutions of these equations have damped oscillatory
behavior. 

Our study demonstrates that dilaton stabilization via gaugino 
condensation and radion stabilization via enhanced symmetry string
states are compatible. Thus, we conclude that in Heterotic string
theory (which admits enhanced symmetry states which are massless
at the enhanced symmetry point), the joint action of string gases
and gaugino condensation provides a means to stabilize all of the
string moduli, and thus provides an alternative to the usual
flux stabilization scenarios.  In fact, we have constructed an explicit
compactification which can realize both of these mechanisms, the $T^6/\Z$
orbifold, and which can also be modified to produce a semi-realistic low
energy spectrum.

\begin{acknowledgments}

We would like to acknowledge interesting conversations with S. Watson.

This work is supported by NSERC through the Discovery Grant
program. RB is also supported in part by the Canada Research Chairs 
program and by funds from a FQRNT Team Grant. 
AF is supported in part by the Institute for Particle Physics and the 
Perimeter Institute.  RD is supported in part by the Chalk-Rowles Fellowship.

\end{acknowledgments}

\appendix

\section{String spectrum on the orbifold}\label{s:spectrum}

We now discuss the spectrum of the heterotic string on the $T^6/\Z$ orbifold,
following \cite{Polchinski:1998rr} for the Kaluza-Klein zero modes. 

\subsection{KK zero modes}\label{s:zeromodes}
The spectrum of perturbative string theory breaks into two sectors, twisted
and untwisted.  The untwisted sector, with which we concern ourselves, 
is just the spectrum on the $T^6$ projected onto those states which are 
invariant under the $\Z$ transformation.  To determine the behavior of 
string states under the $\Z$, we should consider the transformation of the
worldsheet fields.  The bosonic worldsheet fields are just the target space
coordinates, so they transform as in (\ref{Z3one}).  The superconformal
ghost fields are inert under the orbifold transformation.

The remaining worldsheet fields are fermionic.  On the right-moving side
of the string, there are three complex fermions $\tilde\psi$, the 
superpartners of the $Z$s.  By worldsheet supersymmetry, these transform in
the same manner as the bosonic coordinates:
\eq{Z3two}{
\tilde\psi^{1,2}\simeq \alpha \tilde\psi^{1,2}\ ,\ \ 
\tilde\psi^3\simeq \alpha^{-2}\tilde\psi^3\ .}
The left-moving fermions are 16 complex fermions, which are responsible for the
spacetime gauge group.  These transform under the diagonal $\Z$ as
\eq{Z3three}{
\lambda^{1,2}\simeq \alpha \lambda^{1,2}\ ,\ \ 
\lambda^3\simeq \alpha^{-2}\lambda^3\ ,\ \ \lambda^{4,\ldots 16}
\simeq \lambda^{4,\ldots 16}\ .}
We reiterate that these transformations are dictated by our choice of 
behavior for the spacetime gauge symmetry under the orbifold projection.

Each fermion can also take periodic or antiperiodic boundary conditions on the
spatial worldsheet coordinate. The $\tilde\psi$ all have the same 
periodicity (the periodic sector is the
``Ramond'' sector, labeled $\tilde R$, and the antiperiodic is 
``Neveu-Schwarz,'' labeled $\tNS$).  In the $E_8\times E_8$ string,
the two sets of complex fermions $\lambda^1,\ldots \lambda^8$ and 
$\lambda^9,\ldots \lambda^{16}$ have matching periodicities, leading to four
sectors, labeled (as for the $\tilde\psi$) $\NS$-$\NS'$, $R$-$\NS'$, 
$\NS$-$R'$, and $R$-$R'$.  For our purposes, the reader should know that
$R$ sector fermions are integer moded, while $\NS$ sector fermions are 
half-integer moded.  
For more details of the consistent construction of the spectrum,
we refer the reader to \cite{Polchinski:1998rr} and other reviews.

We can now list the states that are invariant under the orbifold projection,
working at the massless level.
We start with spacetime gauge singlets; the bosons are
\eq{sugra-b}{\alpha_{-1}^\mu \t\psi^\nu_{-1/2}\ket{0}\ ,\ \ 
\alpha_{-1}^i\t\psi_{-1/2}^{\bj} \ket{0}\ ,\ \ \alpha_{-1}^{\bi}
\t\psi_{-1/2}^{j} \ket{0}\ .}
As usual, the $\alpha^{\mu,m}_{\pm n}$ are the mode operators for the 
worldsheet bosons.
The first set of states includes the 4D graviton, dilaton, and axion, and
the other two sets give the internal components of the metric and 2-form.
Notice that the internal 
metric is now required to be Hermitean, which restricts the
moduli space somewhat.  In addition, the Kaluza-Klein graviphoton is projected
out.  The spacetime fermions are in the $\t R$ sector,
and the contribution of the $\t\psi$ fermions enters only through their ground
state, which transforms as a chiral spinor in 10D.  These decompose into 4D
spinors and either singlets or triplets under the $SU(3)$ holonomy group; 
the fundamental and antifundamental have $\Z$ eigenvalues $\alpha$ and 
$\alpha^2$ respectively.  Therefore, the gauge singlet fermions are
\eq{sugra-f}{\alpha_{-1}^\mu\ket{s_4,\mathbf{1}}\ ,\ \ 
\alpha_{-1}^i\ket{s_4,\mathbf{3}}\ ,\ \ \textnormal{and conjugates.}}

Now we consider massless states charged under the gauge group.
Before we impose the orbifold projection, it is useful to enumerate the 
possible states of the gauge theory.  Since one $E_8$ factor will be broken
to $SU(3)\times E_6$ by the $\Z$ projection, 
which is the center of the $SU(3)$,
we will classify the states by their representation under 
$SU(3)\times E_6\times E_8$.  At the massless level, we need only consider
the $\NS$-$\NS'$, $R$-$\NS'$, and $\NS$-$R'$ sectors, since the $R$-$R'$ sector
has a positive mass already in the ground state.  In fact, to make a complete
representation of $E_8$, it takes half of the $\NS$-$\NS'$ states and all of
either the $R$-$\NS'$ or $\NS$-$R'$ states.  Under $SU(3)\times E_6\times E_8$,
we have the adjoints
\eq{gauge-adj}{\ket{\m 8,\m 1, \m 1}\ , \ \ \ket{\m 1, \m{78}, \m 1}\ ,\ \ 
\ket{\m 1, \m 1, \m{248}}\ ,}
which are invariant under $\Z$, and the conjugate pair
\eq{gauge-conj}{\ket{\m 3,\m{27},\m 1}\ \textnormal{and}\ 
\ket{\b{\m{3}}, \overline{\m{27}}\, \m 1}\ ,}
which have $\Z$ eigenvalues $\alpha$ and $\alpha^2$ respectively.  Again, 
these eigenvalues follow because our $\Z$ is the center of the $SU(3)$ factor.

With these results, we can assemble $\Z$ invariant states simply.  The 
surviving massless bosons are
\eq{gauge-b}{\t\psi_{-1/2}^\mu \ket{\m{adj}}\ ,\ \ 
\t\psi_{-1/2}^i\ket{\b{\m 3},\overline{\m{27}}\, \m 1}\ , \ \ 
\t\psi_{-1/2}^{\bi} \ket{\m 3,\m{27},\m 1}\ .}
Here, $\m{adj}$ contains the adjoints of all the gauge factors.  The 
spacetime vectors are the gauge bosons for the remaining gauge group, and
the scalars in the latter two states are given 
by components $A_{m}$ of the 10D gauge field.  The massless
fermions on the orbifold are
\eq{gauge-f}{\ket{s_4,\m 1;\m{adj}}\ ,\ \ket{s_4,\b{\m 3};\m 3,\m{27},\m 1} 
\ ,}
and their conjugates.  These are the gaugini and chiral fermions.

Finally, we should note that we have considered only the untwisted sector
of the $T^6/\Z$ spectrum.  There is an additional ``twisted sector'' of 
string states localized around each of the 27 fixed points of the orbifold.
However, these states are unimportant for our concerns, so we do not 
discuss them here.  We refer the reader to, for example, 
\cite{Polchinski:1998rr} for a review of the twisted sectors.

\subsection{Kaluza-Klein spectrum}\label{s:kkspectrum}

Now we turn to the (KK) momentum and winding spectrum.  
Our main interest will be to prove that strings carrying KK momentum and 
winding can become massless at special values of the radius.
Because we have chosen
a factorized torus, we can consider the states of a single $T^2$ at a time.

The first thing we need to know is the quantization of the winding and 
momentum.  Winding is a vector $w^z$, $w^{\b z}$, which we denote $w,\b w$.
It is simply quantized in units of the torus periodicity, so the shortest
units of winding are $w=1,\alpha$ (with length 1).  
An odd feature of the torus is that $w=\alpha^2=-1-\alpha$ also has unit
length, so we can think of winding as being quantized in units of 
$w=1,\alpha,\alpha^2$.  Momentum is somewhat trickier.  Writing $n_z=n$,
$n_{\b z}=\b n$, we require that $(nZ+\b n \b Z)/\sqrt{\alp}$ 
shift by an integer multiple of $2\pi$ when $Z\to Z+2\pi\sqrt{\alp}$ and when
$Z\to Z+2\pi\alpha\sqrt{\alp}$.  With a little work, it is possible to
see that the minimum-length momenta are $n=\hat\alpha\equiv (1-i/\sqrt{3})/2$,
$n=\alpha^2\hat\alpha\equiv \alpha\cdot \hat\alpha$, and 
$n=\alpha\hat\alpha\equiv \alpha^2\cdot\hat \alpha$.  In the latter two 
momenta, the $\equiv$ signs designate the fact that $n$ is a covector, so
a single $\Z$ rotation by $\alpha$ of the coordinates rotates $n$ by 
$\alpha^{-1}=\alpha^2$.

Using the $\Z$ transformation of momentum and winding given above, we see
that the KK $\Z$ eigenstates are quantum superpositions
\eq{KK1}{\left(\ket{n,w}+\alpha^k\ket{\alpha^2 n,\alpha w}
+\alpha^{2k}\ket{\alpha n,\alpha^2 w}\right)}
for $k=0,1,2$ with eigenvalue $\alpha^{2k}$ (ignoring normalization).

Let us now consider the masses associated with KK momentum and winding.  
The general formula in complex coordinates is
\bea
m^2 &=& \frac{4}{\alp}g^{i\bj} \left( n_i +B_{i\b k}\b w^{\b k}
\pm g_{i\b k} \b w^{\b k}\right)\nonumber\\
&&\times\left( \b n_{\bj} +B_{\bj l} w^{l}\pm g_{\bj l} w^{l}\right)+\cdots\ ,
\label{KKmass}\eea
where $\cdots$ represents the contribution from oscillators, which depends
on the fermion periodicity ($R$ vs $\NS$).  The $\pm$ sign is positive for 
the left-moving side of the string and negative for the right-moving side.
Taking the orbifold to factorize into three $T^2/\Z$ orbifolds with diagonal
metric and $B$-field, we find
\eq{KKmass2}{m^2 = \frac{4}{\alp}b^{-2}\left| n+i\frac{1}{2}
\beta \b w \pm \frac{1}{2}b^2\b w\right|^2+\cdots\ .}
Here we treat $n$ and $w$ as three-vectors, but, to start, we will 
limit ourselves to momentum and winding on a single $T^2$ factor.

In the $\NS$-$NS'$-$\widetilde{\NS}$ sector, the mass is given by
\bea 
m^2&=&\frac{4}{\alp}\left[b^{-2}
\left|n-\frac{1}{2}\left(b^2-i\beta\right)\b w\right|^2
+ \left(\t N-\frac{1}{2}\right)\right]\nonumber\\
&=&\frac{4}{\alp}\left[b^{-2}
\left|n+\frac{1}{2}\left(b^2+i\beta\right)\b w\right|^2
+(N\!-\!1)\right]\!\! .\label{KK2}
\eea
Here $N$ and $\t N$ are the left- and right-moving oscillator excitation 
number of the string; $N$ must be integer, but $\t N>0$ may be half-integer.  
Note that the second equality implies the 
``level-matching'' constraint
\eq{level1}{ n \cdot w + \b n\cdot \b w =\t N +\frac{1}{2}-N \ .}
It is clear that these states are massive for generic values of the
moduli $b,\beta$.  However, at special points in moduli space, some KK 
excitations can become massless.  For example, at $b=1,\beta=1/\sqrt{3}$, 
we can take $n=\pm \hat\alpha$, $w=\pm 1$,
$N=0$, and $\t N =1/2$ to get the massless states
\bea
&&\t\psi^\mu_{-1/2}\left( \ket{\hat\alpha,1}+\ket{\alpha^2\hat\alpha, \alpha }
+\ket{\alpha\hat\alpha ,\alpha^2}\right)\nonumber\\
&&\t\psi^{i}_{-1/2}\left( \ket{\hat\alpha,1}+\alpha\ket{\alpha^2\hat\alpha,
\alpha}+\alpha^2\ket{\alpha\hat\alpha,\alpha^2}\right)\nonumber\\
&&\t\psi^{\bi}_{-1/2}\left( \ket{\hat\alpha,1}+\alpha^2\ket{\alpha^2\hat\alpha
,\alpha}+\alpha\ket{\alpha\hat\alpha,\alpha^2}\right)\label{enhanced1}\eea
(and states with $n,w\to -n,w$).
By supersymmetry, there are also massless states in the $\NS$-$\NS'$-$\t R$
sector at $b=1$:
\bea
&&\ket{s_4,\m 1}\otimes 
\left( \ket{\hat\alpha,1}+\ket{\alpha^2\hat\alpha, \alpha }
+\ket{\alpha\hat\alpha ,\alpha^2}\right)\nonumber\\
&&\ket{s_4,\b{\m 1}}\otimes 
\left( \ket{\hat\alpha,1}+\ket{\alpha^2\hat\alpha, \alpha }
+\ket{\alpha\hat\alpha ,\alpha^2}\right)\nonumber\\
&&\ket{s_4,\m 3}\otimes 
\left( \ket{\hat\alpha,1}+\alpha\ket{\alpha^2\hat\alpha,
\alpha}+\alpha^2\ket{\alpha\hat\alpha,\alpha^2}\right)\nonumber\\
&&\ket{s_4,\b{\m 3}}\otimes 
\left( \ket{\hat\alpha,1}+\alpha^2\ket{\alpha^2\hat\alpha
,\alpha}+\alpha\ket{\alpha\hat\alpha,\alpha^2}\right)\! .\label{enhanced2}\eea
Near the enhanced symmetry point ($b=1+\delta b$, $\beta=1/\sqrt{3}$), 
these states have $m^2=4\delta b^2/\alp$.  

In fact, the states (\ref{enhanced1},\ref{enhanced2}) are all the possible
extra massless states in this sector (because consistency of the string
theory requires $\t N\geq 1/2$), up to some permutations.  Taking 
$n=\alpha^2\hat\alpha$ and $w=\alpha$ or $\alpha\hat\alpha$ and $\alpha^2$ 
would be acceptable, but these are identical to the states
we have already presented by the orbifold projection.  Therefore, they are
not distinct physical states.  The only way to get distinct physical states
is to note that we really have three $T^2$ factors, so $n$ and $w$ are 
complex 3-vectors.  The mass (\ref{KK2}) and level-matching condition
(\ref{level1}) still apply as written.  However, getting 
$n\cdot w + \b n\cdot \b w =1$ still requires all the momentum and winding to
lie on the same $T^2$ factor.  Therefore, we get six copies of the states
(\ref{enhanced1},\ref{enhanced2}), two for each $T^2$ factor.  Because of
the extra vector particles, the spacetime gauge group gets an extra factor
of $U(1)^2$ for each $T^2$ factor.  Hence, $T=1+i/\sqrt{3}$ 
is called an ``enhanced symmetry point'' of the compactification.

We can actually see that the other sectors do \emph{not}
have masseless states at $T=1+i/\sqrt{3}$.  
The $R$-$R'$ sector is already massive in
the ground state, so it can have no massless states.  Furthermore, the 
$\NS$-$R'$-$\widetilde{\NS}$ and $R$-$\NS'$-$\widetilde{\NS}$ 
states have mass-shell relations
\bea 
m^2&=&\frac{4}{\alp}\left[b^{-2}
\left|n-\frac{1}{2}\left(b^2-i\beta\right)\b w\right|^2
+ \left(\t N-\frac{1}{2}\right)\right]\nonumber\\
&=&\frac{4}{\alp}\left[b^{-2}
\left|n+\frac{1}{2}\left(b^2+i\beta\right)\b w\right|^2
+N\right] .\label{KK3}
\eea
with level-matching condition
\eq{level2}{n\cdot w + \b n\cdot \b w =\t N -\frac{1}{2}-N \ .}
The only possibility for extra massless states with $\t N>0$ is to take 
$\t N=1/2$ and $N=0$.  However, we clearly see then that $n=w=0$.  Therefore,
we have listed all the possible extra states at the enhanced symmetry point.

Let us also address a concern
raised in \cite{Patil2}.  The authors of \cite{Patil2}
studied the bosonic string and found states with masses $m^2\propto \delta b$
for $b=1+\delta b$ in addition to the type of states we have found.  These
strings become tachyonic for some values of the radius, and their absence
in our case is related to supersymmetry.  In particular, we are required 
to take $\t N\geq 1/2$ to formulate a consistent supersymmetric string theory,
and it is precisely this condition that has limited our possibilities so 
much.

For later reference, let us rewrite the mass formula (\ref{KK2}) for the
special string states in terms of 10D Einstein frame variables.  The 
10D metrics are related by the transformation
\eq{stringEinstein}{g^E_{mn} = e^{-\phi/2}g^s_{mn}\ , }
where $\phi$ is the 10D dilaton.
(and similarly for the external components).  This transformation law means
that we should replace $b\to e^{\phi/4}b$ to get the mass formula
\eq{KKfinal2}{
m^2 = \frac{1}{\alp} \left(e^{-\phi/4}b^{-1}-e^{\phi/4}b\right)^2 
\ .}  We will use this version from now on.

\bibliography{stable}

\end{document}